\documentclass [11pt] {article}
\usepackage{latexsym,bbold,epsf,colordvi,cite,colordvi}
\newcommand{\slashed}[1]{\rlap{$#1$}/}
\newcommand{\slashp}{\mbox{$\not \hspace*{-1.10mm} p$}}

\newcommand{\GeV}{\mbox{\rm GeV}}

\newcommand{\lsim}[1]{
\setlength{\unitlength}{12pt}
\begin{picture}(1.4,1.)
\put(.7,-0.3){\makebox(0.0,1.)[t]{$<$}}
\put(.7,-0.3){\makebox(0.0,1.)[b]{$\sim$}}
\end{picture}#1}

\newcommand{\eff}{\mbox{\rm\scriptsize eff}}
\newcommand{\pert}{\mbox{\rm\scriptsize pert}}
\newcommand{\Npert}{\mbox{\rm\scriptsize Npert}}

\begin{document}

%\draft

%\preprint{ day. Month. 2003. }

%\vskip 1.5cm

\title{
A Bethe-Salpeter study with the $\langle A^2 \rangle$-enhanced
effective QCD coupling 
}

\author{Dalibor Kekez$^a$ and Dubravko Klabu\v{c}ar$^b$ 
\and
$^a${\footnotesize Rudjer Bo\v{s}kovi\'{c} Institute,
         P.O.B. 180, 10002 Zagreb, Croatia} \\
$^b${\footnotesize Department of Physics, Faculty of Science, Zagreb University} \\
{\footnotesize Bijeni\v cka c. 32, 10000 Zagreb, Croatia}}

\date{}

\maketitle

\vspace*{-10mm}
\begin{abstract}

\noindent 
Dyson-Schwinger equations provide a prominent 
approach to physics of strong interactions. To reproduce 
the hadronic phenomenology well, the Dyson-Schwinger 
approach in the rainbow-ladder approximation must employ 
an effective interaction between quarks which is fairly 
strong at intermediate ($Q^2 \sim 0.5$ GeV$^2$) spacelike
transferred momenta. We have recently proposed that such
an interaction may originate from the dimension 2 gluon 
condensate $\langle A^2 \rangle$ which has recently 
attracted much attention, and showed
that the resulting effective running coupling leads
to the sufficiently strong dynamical chiral symmetry
breaking and successful phenomenology at least in the
light sector of pseudoscalar mesons. In the present
paper, we give a more detailed investigation of the
parameter dependence of these results.

\end{abstract}
%\pacs
{ Pacs: 11.10.St; 11.30.Qc; 12.38.Lg; 14.40.Aq }

\section{Introduction}
\label{INTRO}

In recent years, the dimension 2 gluon condensate
$\langle A_\mu^a A^{a\mu}\rangle \equiv \langle A^2 \rangle$
attracted a lot of theoretical attention
\cite{Boucaud:2000nd,Gubarev:2000eu,Gubarev:2000nz,Kondo:2001nq,Kondo:2001tm,Boucaud:2002fx,Dudal:2002xe},
to quote just several of many papers
offering evidence that this condensate may be important 
for the nonperturbative regime of Yang-Mills theories, 
particularly QCD. Although $\langle A^2 \rangle$ is 
not gauge invariant, it was even argued that its value 
in the Landau gauge may have a physical meaning
\cite{Gubarev:2000eu,Gubarev:2000nz,Dudal:2002xe}.
In our recent paper \cite{Kekez:2003ri} we argued that 
$\langle A^2 \rangle$ may be relevant for the Dyson-Schwinger
(DS) approach to QCD. Namely, in order that this approach 
leads to
a successful hadronic phenomenology, an enhancement of
the effective quark-gluon interaction seems to be needed
at intermediate ($Q^2 \sim 0.5$ GeV$^2$) momenta{\footnote{{We 
adopt the convention $k^2 = -Q^2 < 0$ for spacelike momenta $k$.}}},
and Ref. \cite{Kekez:2003ri} showed that the gluon 
condensate $\langle A^2 \rangle$ provides such an 
enhancement. It also showed that 
the resulting effective strong running coupling leads
to the sufficiently strong dynamical chiral symmetry
breaking and successful phenomenology in the light 
sector of pseudoscalar mesons. However, the issue
of the parameter dependence of the results was
just commented on very briefly. 
Thus, in the present paper, 
in Sec. \ref{alpha_effPhenomenology},
we give a more detailed investigation and presentation 
of the parameter dependence of these results.
A brief recapitulation of the DS approach and the
effective interaction it needs is given in the next section.

\section{DS approach and its effective interaction}
\label{DSapp}

DS approach to hadrons and their quark-gluon
substructure \cite{Alkofer:2000wg,Maris:2003vk,Roberts:2003kd}
has strong and clear connections with QCD. Besides being covariant,
this approach is chirally well-behaved and nonperturbative. This
has been crucial, especially in the light-quark sector of QCD,
for successful descriptions of bound states achieved by 
phenomenological DS studies (e.g., see recent reviews 
\cite{Maris:2003vk,Roberts:2003kd} and references therein),
where one can treat soundly even the processes influenced by axial 
anomaly{\footnote{See, e.g., Refs. \cite{Maris:1998hc,Roberts:1994hh}
for the $\pi^0 \to \gamma\gamma$ transition amplitude 
$T_{\pi^0}^{\gamma\gamma}$, and Refs. 
\cite{Alkofer:1995jx,Bistrovic:1999yy,Bistrovic:1999dy,Cotanch:2003xv}
for the related transition $\gamma \to \pi^+\pi^0\pi^-$.}}, 
which is really remarkable for a bound-state approach.
What happens is that in the process of solving DS equations,
one in essence derives a constituent quark model which turns
out to be successful over a very wide range of masses. Its
chief virtue is that it incorporates the correct chiral symmetry
behavior through the gap equation for the full, dynamically 
dressed quark propagator $S_q$ and the Bethe-Salpeter (BS) equation 
for the bound states of the dynamically dressed quarks (and
antiquarks). That is, the constituent quarks arise through
dressing resulting from dynamical chiral symmetry breaking 
(D$\chi$SB) in the (``gap'') DS equation for the full quark 
propagators, while the {\it light} $q\bar q$ pseudoscalar 
solutions of the BS equation (in a {\it consistent} approximation) 
are ({\it almost} massless) {\it quasi}-Goldstone bosons of 
D$\chi$SB.  Generation of D$\chi$SB is well-understood 
\cite{Alkofer:2000wg,Maris:2003vk,Jain:1991pk,Munczek:1991jb,Jain:qh,Maris:1997tm,Maris:1999nt}
in the rainbow-ladder approximation (RLA). Thus, phenomenological 
DS studies have mostly been relying on RLA and using 
{\it Ans\" atze} of the form
\begin{equation}
[K(k)]_{ef}^{hg} = i 4\pi\alpha_{\mbox{\rm\scriptsize eff}}(-k^2) \,
       D_{\mu\nu}^{ab}(k)_{0} \,
[\frac{\lambda^a}{2}\,\gamma^{\mu}]_{eg}
[\frac{\lambda^b}{2}\,\gamma^{\nu}]_{hf}
\label{RLAkernel}
\end{equation}
for interactions between quarks. In this equation,
$e,f,g,h$ schematically represent spinor, color and flavor 
indices and $D_{\mu\nu}^{ab}(k)_{0}$ is the 
{\it free} gluon propagator in the gauge in which the 
aforementioned DS studies have been carried out almost 
exclusively, namely the Landau gauge:
\begin{equation}
D_{\mu\nu}^{ab}(k)_{0} = \frac{\delta^{ab}}{k^2}
(-g_{\mu\nu} + \frac{k_\mu k_\nu}{k^2}) \, ,
\end{equation}
while $\alpha_{\mbox{\rm\scriptsize eff}}(Q^2)$ is an effective 
running coupling on which we will comment below at length.

\begin{figure}
\begin{center}
\epsfxsize = 12 cm \epsfbox{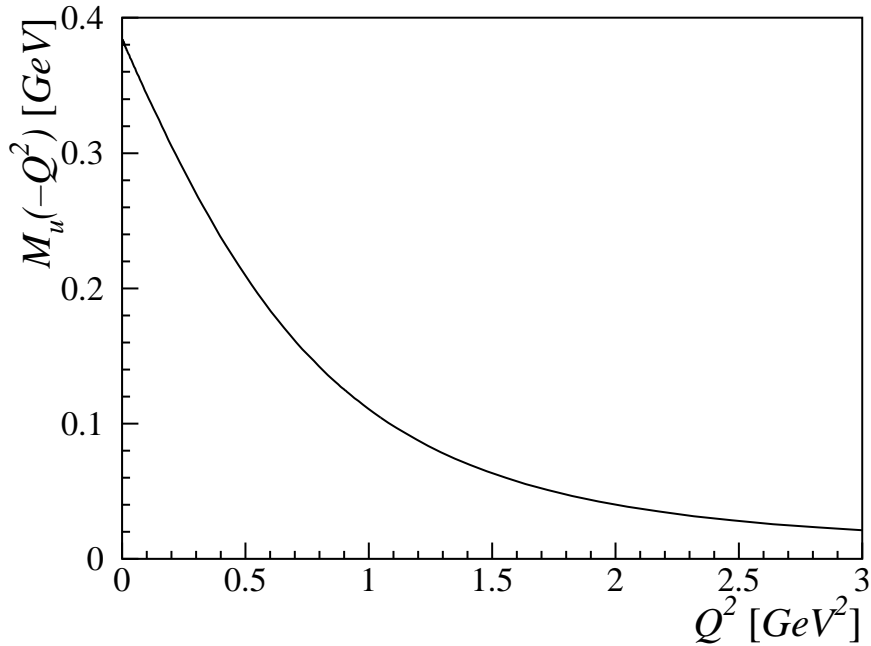}
\end{center}
\caption{The effective non-strange ($q=u$) quark mass function 
$M_u(-Q^2)$ calculated using the effective coupling (\ref{ourAlpha_eff}) 
proposed in Ref. \cite{Kekez:2003ri} and the input parameters given 
by Eqs. (\ref{StandardParameterSet-new}) and (\ref{newBAREmValues}).
        }
\label{fig:Mx-lin}
\end{figure}

The BS equation for the bound-state vertex 
$\Gamma_{q{\bar q}'}$ of the meson composed of the quark 
of the flavor $q$ and antiquark of the flavor $q'$, is then
\begin{equation}
[\Gamma_{q{\bar q}'}(k,{P})]_{ef} = \int \frac{d^4\ell}{(2\pi)^4} 
[S_q(\ell+\frac{{P}}{2}) \Gamma_{q{\bar q}'}(\ell,{P}) 
S_{q'}(\ell-\frac{{P}}{2}) ]_{gh} [K(k-\ell)]_{ef}^{hg}~.
\label{BSE}
\end{equation}
The consistent RLA requires that the same interaction kernel
(\ref{RLAkernel}) be previously used in the DS equation for 
the full quark propagator $S_q$.
That is, 
dressed quark propagators $S_q(k)$ for various flavors $q$,
        \begin{equation}
 S_q^{-1}(p)
        =
        A_q(p^2)\slashed{p} - B_q(p^2)~, \qquad (q=u,d,s,...)~,
        \label{quark_propagator}
        \end{equation}
%[to be contrasted with the free quark propagators
%$S_q(k)_0 = (\slashk - \widetilde{m}_q)^{-1}$] 
are obtained by solving the gap DS equation
\begin{equation}
        S_q^{-1}(p) = \slashp - \widetilde{m}_q - \,
i 4\pi \int \!\frac{d^4\ell}{(2\pi)^4} \, \alpha_{\eff}(-(p-\ell)^2)
D_{\mu\nu}^{ab}(p-\ell)_{0} \, \frac{\lambda^a}{2}\,\gamma^{\mu} 
S_q(\ell) \frac{\lambda^b}{2}\,\gamma^{\nu}~.
       \label{SD-equation}
        \end{equation}
Following the approach of Munczek and Jain \cite{Munczek:1991jb,Jain:qh}, 
the gap equation (\ref{SD-equation}) is unrenormalized, but regularized 
by an ultra-violet cutoff $L$. This cutoff is however huge compared to 
the QCD scale $\Lambda_{QCD}$. 
(In the present paper, $L = 134$ GeV as in Ref. \cite{Jain:qh}.)
In Eq. (\ref{SD-equation}), $\widetilde{m}_q$ is the cutoff-dependent 
bare mass of the quark flavor $q$ breaking the chiral symmetry explicitly. 
The case $\widetilde{m}_q=0$ corresponds to the chiral limit where
the current quark mass $m_q=0$, and where the constituent quark mass
$M_q(0) \equiv B_q(0)/A_q(0)$ stems exclusively from 
the {\it nonperturbative} phenomenon of D$\chi$SB.
Of course, calling the ``constituent  mass" the value of the
``momentum-dependent constituent mass function" 
$M_q(p^2) \equiv B_q(p^2)/A_q(p^2)$
at exactly {$p^2=0$} and not at some other low {$-p^2$}, is a matter
of a somewhat arbitrary choice. However, it is just a matter of 
terminology and nothing essential.
What is important to get a successful hadronic phenomenology,
especially in the light-quark sector $(q=u,d,s)$, is that 
D$\chi$SB is sufficiently strong. This means that the gap equation
(\ref{SD-equation}) should yield quark propagator solutions
$A_q(p^2)$ and $B_q(p^2)$ giving the dressed-quark mass function 
$M_q(p^2)$ whose values {\it at low} {$-p^2$} are of the order of 
typical constituent mass values, namely several hundred MeV, 
even in the chiral limit. A typical example of such $M_q(p^2)$
is given in Fig. \ref{fig:Mx-lin}, obtained with 
$\alpha_{\mbox{\rm\scriptsize eff}}(Q^2)$ (\ref{ourAlpha_eff})
proposed originally in our Ref. \cite{Kekez:2003ri} and 
further advocated in the present paper.

\begin{figure}
%\centerline{\includegraphics[height=67mm,angle=0]{alphas.eps}}
\begin{center}
\epsfxsize = 14 cm \epsfbox{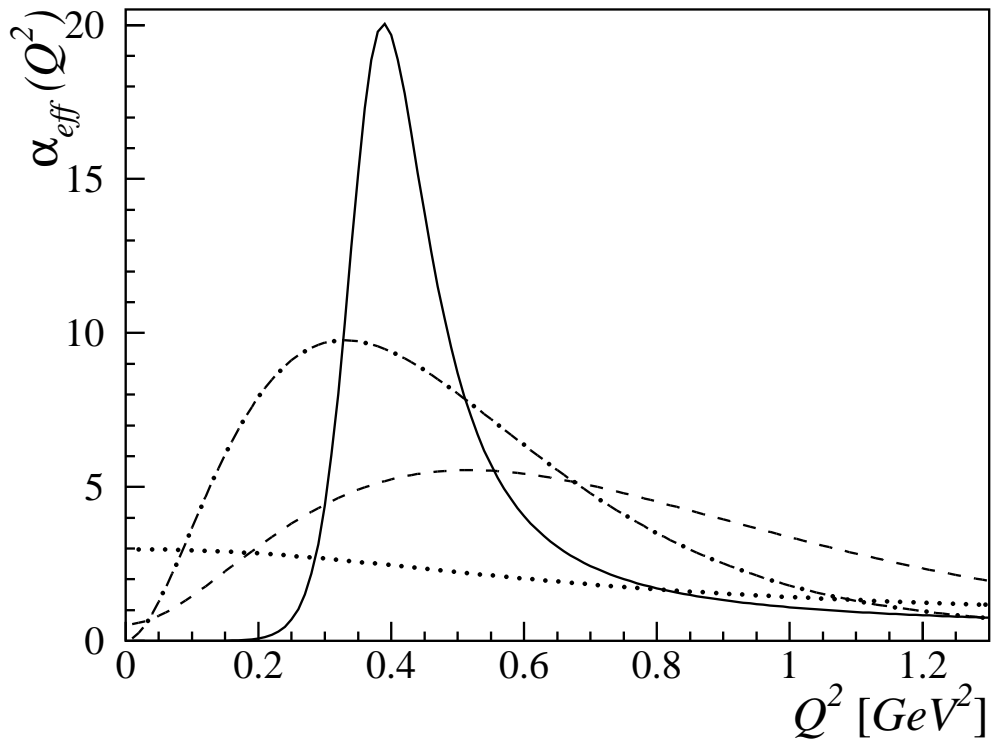}
\end{center}
\caption{The momentum dependence of various strong running couplings
mentioned in the text. JM \cite{Jain:qh} and MRT
\cite{Maris:1999nt,Maris:2003vk} $\alpha_{\eff}(Q^2)$
are depicted by, respectively, dashed and dash-dotted curves.
The effective coupling (\ref{ourAlpha_eff}) proposed and analyzed
in the present paper is depicted by the solid curve, and
$\alpha_{\mbox{\rm\scriptsize s}}(Q^2)$~(\ref{Alkofalpha})
of Fischer and Alkofer \cite{Fischer:2003rp} (their fit A)
by the dotted curve.}
\label{Fig1}
\end{figure}

Indeed, the issue of the origin of the interaction 
(\ref{RLAkernel}), or, equivalently, 
$\alpha_{\mbox{\rm\scriptsize eff}}(Q^2)$ which would enable
successful phenomenology is crucial for the DS studies.
The form of $\alpha_{\eff}$ is only partially known from the fact that 
at large spacelike momenta it must reduce to $\alpha_{\pert}(Q^2)$, 
the well-known running coupling of perturbative QCD.
% (PQCD).
However, for momenta $Q^2 \lsim 1$ GeV$^2$, where non-perturbative QCD 
%(NPQCD) 
applies, the interactions are still not known; therefore, 
in phenomenological DS studies, $\alpha_{\eff}(Q^2)$ must 
be modeled for $Q^2 \lsim 1$ GeV$^2$ - e.g., see Refs.
\cite{Munczek:1983dx,Jain:qh,Maris:1997tm,Maris:1999nt,Alkofer:2000wg,Maris:2003vk,Roberts:2003kd}.
%and references therein. 
There, one can see that phenomenologically most successful of those 
modeled interactions have a rather large bump at the intermediate 
momenta, around $Q^2 \sim 0.5$ GeV$^2$. 
For example, in Fig. \ref{Fig1} compare $\alpha_{\eff}(Q^2)$ used 
by Jain and Munczek (JM) \cite{Jain:qh} and by Maris, Roberts and Tandy 
(MRT) \cite{Maris:1997tm,Maris:1999nt,Maris:2003vk,Roberts:2003kd}.
In any case, 
successful DS phenomenology requires that this modeled part of 
the interaction (\ref{RLAkernel}) be fairly strong.  That is, 
regardless of details of the interaction, its 
{\it integrated strength} in the infrared must be fairly high 
to achieve acceptable description of hadrons, notably mass spectra 
and D$\chi$SB \cite{Maris:2003vk,Roberts:2003kd}. 

Theoretical explanations on what could be the origin of so strong 
nonperturbative part of the phenomenologically required interaction 
are obviously very much needed, either from the {\it ab initio} 
studies of sets of DS equations for Green's functions of QCD 
(see, e.g., the recent review \cite{Alkofer:2000wg})
or from somewhere outside DS approach. 
The particularly important result of the {\it ab initio} DS studies
is that, in the Landau gauge, the effects of ghosts are absolutely
crucial for the intermediate-momenta enhancement of the effective 
quark-gluon interaction 
\cite{Alkofer:2000wg,Alkofer:2002ne,Fischer:2003rp,Alkofer:2003vj,Bloch:2003yu}. 
This is obvious in the expression for the strong running coupling 
$\alpha_{\mbox{\rm\scriptsize s}}(Q^2)$ in these Landau-gauge studies 
\cite{Alkofer:2000wg,Alkofer:2002ne,Fischer:2003rp,Alkofer:2003vj,Bloch:2003yu},
\begin{equation}
\alpha_{\mbox{\rm\scriptsize s}}(Q^2)
=
\alpha_{\mbox{\rm\scriptsize s}}(\mu^2) \, Z(Q^2) \, G(Q^2)^2 
\, ,
\label{Alkofalpha}
\end{equation}
where $\alpha_{\mbox{\rm\scriptsize s}}(\mu^2) = g^2/4\pi$ and 
$Z(\mu^2) G(\mu^2)^2 = 1$ at the renormalization point $Q^2 = \mu^2$.
The gluon renormalization function $Z(-k^2)$ defines the full gluon 
propagator $D_{\mu\nu}^{ab}(k)$ in the Landau gauge:
\begin{equation}
D_{\mu\nu}^{ab}(k) = Z(-k^2) D_{\mu\nu}^{ab}(k)_{0} = \frac{Z(-k^2)}{k^2} \,\,
\delta^{ab} \left( -g_{\mu\nu} + \frac{k_\mu k_\nu}{k^2} \right)
\, .
\label{gluonLGpropag}
\end{equation}
Similarly, $G(-k^2)$ is the ghost renormalization function 
which defines the full ghost propagator 
$D_G^{ab}(k) = \delta^{ab} G(-k^2)/k^2$.

While the {\it ab initio} DS studies 
\cite{Alkofer:2000wg,Alkofer:2002ne,Fischer:2003rp,Alkofer:2003vj,Bloch:2003yu} 
do find significant enhancement of $\alpha_{\mbox{\rm\scriptsize s}}(Q^2)$, 
Eq. (\ref{Alkofalpha}), until recently this seemed still not enough 
to yield a sufficiently strong D$\chi$SB 
(e.g., see Sec. 5.3 in Ref. \cite{Alkofer:2000wg})
and a successful phenomenology. However, for carefully constructed 
dressed quark-gluon vertex {\it Ans\" atze},
Fischer and Alkofer \cite{Fischer:2003rp} have recently 
managed to obtain good results for dynamically generated 
constituent quark masses and pion decay constant $f_\pi$,
although not simultaneously also for the chiral quark-antiquark
$\langle \bar{q}q \rangle$ condensate, which then came out 
somewhat larger than the phenomenological value.  Thus, 
the overall situation is that there is progress in this direction 
\cite{Alkofer:2002ne,Fischer:2003rp,Alkofer:2003vj,Bloch:2003yu,Maris:2002mt}, 
but that further investigation and elucidation of 
the origin of phenomenologically successful effective interaction 
kernels remains one of primary challenges in contemporary DS studies
\cite{Maris:2003vk,Roberts:2003kd}. This provided the motivation
for our paper \cite{Kekez:2003ri}, where we pointed out that such 
an interaction kernel for DS studies in the Landau gauge resulted 
from cross-fertilization of the DS ideas on the running coupling 
of the form (\ref{Alkofalpha}) 
\cite{Alkofer:2000wg,Alkofer:2002ne,Fischer:2003rp,Alkofer:2003vj,Bloch:2003yu} 
and the ideas on the possible relevance of the dimension 2 gluon
condensate 
$\langle A_\mu^a A^{a\mu} \rangle \equiv \langle A^2 \rangle$
\cite{Boucaud:2000nd,Gubarev:2000eu,Gubarev:2000nz,Kondo:2001nq,Kondo:2001tm,Boucaud:2002fx,Dudal:2002xe,Lavelle:eg,Lavelle:xg,Ahlbach:ws,Lavelle:yh}. 

In Ref. \cite{Kekez:2003ri}, we gave arguments that the
$\langle A^2 \rangle$-contributions to the OPE-improved gluon ($A$) 
and ghost ($G$) polarization functions (found a long time ago by 
Refs. \cite{Lavelle:eg,Lavelle:xg,Ahlbach:ws,Lavelle:yh} and
more recently confirmed by Kondo \cite{Kondo:2001nq}) 
lead to an effective coupling $\alpha_{\eff}(Q^2)$ given by
\begin{equation}
\alpha_{\eff}(Q^2)
 =
\alpha_{\mbox{\rm\scriptsize pert}}(Q^2) \,
Z^{\Npert}(Q^2) \, G^{\Npert}(Q^2)^2~,
\label{ourAlpha_eff}
\end{equation}
where $\alpha_{\pert}(Q^2)$ is the running coupling 
of perturbative QCD, and 
\begin{eqnarray}
Z^{\Npert}(Q^2)&=&\frac{1}{1
                   +\frac{m_A^2 }{Q^2}
                   + \frac{C_A}{Q^4}}~,
\label{ZOPE}
\\
G^{\Npert}(Q^2)&=&\frac{1}{1
                   -\frac{m_A^2}{Q^2}
                   + \frac{C_G}{Q^4}}~.
\label{ZGOPE}
\end{eqnarray}
The functions $Z^{\Npert}(Q^2)$ and $G^{\Npert}(Q^2)$ are
the nonperturbative ($\Npert$) parts of the, respectively,
gluon and ghost renormalization functions $Z(Q^2)$ and 
$G(Q^2)$. They crucially depend on the quantity $m_A$
which can be interpreted as a dynamically generated
effective gluon mass, and which is proportional to 
the dimension 2 gluon condensate $ \langle A^2 \rangle$.
Concretely, for the Landau gauge (to which we stick 
throughout this paper), the number of QCD colors $N_c = 3$ 
and the number of space-time dimensions $D=4$,
\begin{equation}
m_A^2 = \frac{3}{32} \,\,  g^2 \langle A^2 \rangle = 
        - m_G^2~,
\label{gluonMass}
\end{equation}
where $m_G$ is a dynamically generated effective ghost 
mass. (In a subsequent work, Kondo {\it et al.} \cite{Kondo:2001tm} also
worked out logarithmic corrections to Eq. (\ref{gluonMass})
thanks to which the dynamical gluon mass (and ghost mass) vanishes
as $Q^2 \rightarrow \infty$, as it must according to, e.g.,
Cornwall \cite{Cornwall:1981zr,Cornwall:1998ef}.
However, taking this into account is not necessary at the degree of
refinement and precision at which we work in this paper.)

For $g^2 \langle A^2 \rangle$, the Landau-gauge lattice studies of
Boucaud {\it et al.} \cite{Boucaud:2000nd} yield the value $2.76$ GeV$^2$.
This is compatible with the bound resulting from the discussions of
Gubarev {\it et al.} \cite{Gubarev:2000eu,Gubarev:2000nz} on the
physical meaning of $\langle A^2 \rangle$ (although it is gauge-variant)
and its possible importance for confinement. We thus use this value
in Eq. (\ref{gluonMass}) and obtain
\begin{equation}
m_A = 0.845 \, \, \, {\rm GeV} \, .
\label{mAestimate}
\end{equation}
In our considerations below, this value will turn out to be a
remarkably good initial estimate for the dynamical masses $m_A$
and $m_G$.

The coefficients $C_A$ and $C_G$ appearing in 
$Z^{\Npert}(Q^2)$ (\ref{ZOPE}) and $G^{\Npert}(Q^2)$ (\ref{ZGOPE}),
can, in principle, be related to various other condensates
\cite{Lavelle:xg,Ahlbach:ws,Lavelle:yh},
but some of them are completely unknown at present. 
Therefore, both $C_A$ and $C_G$ should at this point be treated 
as free parameters to be fixed by phenomenology.
Fortunately, Ref. \cite{Kekez:2003ri} managed to make the estimate 
$C_A = (0.640 \,\, {\rm GeV})^4$.
This estimate \cite{Kekez:2003ri} is based on the role of only 
one condensate \cite{Lavelle:ve},
the well-known gauge-invariant dimension 4 condensate 
$\langle F^2 \rangle$ \cite{Shifman:bx}, and thus
misses some (unknown) three- and
four-gluon contributions \cite{Ahlbach:ws,Lavelle:yh}. 
Therefore, and since 
the true value of $\langle F^2 \rangle$ is still 
rather uncertain \cite{Ioffe:2002be}, 
we do not attach too much importance to the above
precise value of $C_A$ but just use it as an inspired 
initial estimate.

There is no similar estimate for $C_G$,
but one may suppose that it would not differ from $C_A$ by orders of
magnitude. We thus try
\begin{equation}
C_G = C_A = (0.640 \,\, {\rm GeV})^4
\label{CGCAestimate}
\end{equation}
as an initial guess. It turns out {\it a posteriori} that
this value of $C_G$ leads to a very good fit to phenomenology.

As we discussed in Ref. \cite{Kekez:2003ri}, Eq. (\ref{ourAlpha_eff}) 
can be justified for relatively high $Q^2$, but not for low $Q^2$. 
For example, $\alpha_{\mbox{\rm\scriptsize pert}}(Q^2)$ must ultimately 
hit the Landau pole as $Q^2$ gets lowered. However, this can be handled 
as in other phenomenological DS studies. Their 
various choices of $\alpha_{\eff}(Q^2)$ usually also contain 
$\alpha_{\mbox{\rm\scriptsize pert}}(Q^2)$, but since handling
the Landau pole problem at the fundamental level is out of their 
scope, they 
\cite{Jain:qh,Maris:1997tm,Maris:1999nt,Kekez:1996az,Klabucar:1997zi,Kekez:1998xr}
just shift{\footnote{As pointed out already by, e.g., Cornwall
\cite{Cornwall:1981zr}, dynamically generated gluon mass can provide
the physical reason for such a change in the arguments of logarithms.
That is, $x_0 \propto m_A^2/\Lambda_{QCD}^2 \sim 10$.}} 
the Landau pole to the timelike momenta in all logarithms appearing
here:
$\ln(Q^2/\Lambda_{QCD}^2) \rightarrow \ln(x_0 + Q^2/\Lambda_{QCD}^2)$.
Presently, we adopt this latter procedure. Concretely, for 
$\alpha_{\mbox{\rm\scriptsize pert}}(Q^2)$ we use throughout the 
{$\overline{\mbox{\rm MS}}$}-scheme two-loop expression used before 
by JM \cite{Jain:qh} and our earlier phenomenological DS studies 
\cite{Kekez:1996az,Klabucar:1997zi,Kekez:1998xr,Kekez:1998rw,Kekez:2000aw}. 
This means we use throughout the infrared (IR) regulator $x_0 = 10$ 
(to which all results are almost totally insensitive), the number 
of quark flavors $N_f = 5$, and $\Lambda_{QCD} = 0.228$ GeV.
These parameters of $\alpha_{\mbox{\rm\scriptsize pert}}(Q^2)$ 
are thereby fixed and do not belong among variable parameters
such as $C_A, C_G$, the variation of which is discussed below.

In the present context, the more serious objection 
to our $\alpha_{\eff}$ (\ref{ourAlpha_eff}) 
is that we cannot in advance give an argument that 
the factor $Z^{\Npert}(Q^2) \, G^{\Npert}(Q^2)^2$ 
in the proposed $\alpha_{\eff}(Q^2)$ (\ref{ourAlpha_eff}) indeed 
approximates well nonperturbative contributions at low $Q^2$
(say, $Q^2 < 1$ GeV$^2$), but can only hope that our results 
to be calculated will provide an {\it a posteriori} 
justification for using it as low as $Q^2 \sim 0.3$ GeV$^2$
[since Eq. (\ref{ourAlpha_eff}) takes appreciable values 
down to about $Q^2 \sim 0.3$ GeV$^2$].
Of course, $Z^{\Npert}(Q^2)$ and $G^{\Npert}(Q^2)$ must 
be wrong in the limit $Q^2 \rightarrow 0$, as they are based 
on the results derived by OPE
\cite{Lavelle:eg,Lavelle:xg,Ahlbach:ws,Lavelle:yh,Kondo:2001nq},
which certainly fail in that limit. For example, detailed 
investigations of the $Q^2 \rightarrow 0$ asymptotic behavior 
in {\it ab initio} DS studies 
\cite{Alkofer:2000wg,Alkofer:2002ne,Fischer:2003rp,Alkofer:2003vj,Bloch:2003yu},
settled down to the conclusion that 
$\alpha_{\mbox{\rm\scriptsize s}}(Q^2)$ remains finite as 
$Q^2 \rightarrow 0$, which is also supported by several lattice calculations
{ \cite{Bloch:2002we,Bloch:2003sk}. }
On the other hand, if the presently interesting $\langle A^2 \rangle$ 
condensate is explained by an instanton liquid, the coupling vanishes as 
$\alpha_{\mbox{\rm\scriptsize s}}(Q^2) \propto Q^4$ \cite{Boucaud:2002fx}, 
which is closer 
to the behavior of our $\alpha_{\eff}(Q^2)$ (\ref{ourAlpha_eff}).
Still, Eqs. (\ref{ZGOPE}) enforce, for small $Q^2$, even much more
dramatic suppression of our $\alpha_{\eff}(Q^2)$ (\ref{ourAlpha_eff}), 
which vanishes as $Q^{12}$. This is an unrealistic artefact of 
the proposed form (\ref{ourAlpha_eff}) when applied down to 
the $Q^2 \rightarrow 0$ limit. Nevertheless, 
because of the integration measure in the integral equations
in DS calculations, integrands at these small $Q^2$ [where our 
$\alpha_{\eff}(Q^2)$ (\ref{ourAlpha_eff}) is doubtlessly too suppressed] 
do not contribute much, at least not to the quantities (such as 
$\langle {\bar q}q \rangle$ condensate, meson masses, decay 
constants and amplitudes) calculated in phenomenological DS analyses.
Hence, the form of $\alpha_{\eff}(Q^2)$ at $Q^2$ close to zero 
is not very important{\footnote{Of course, the $Q^2 \to 0$ domain 
would give an important contribution in a case with a sufficiently 
strong (but still integrable) divergence in $\alpha_{\eff}(Q^2)$, 
such as the delta function in Ref. \cite{Munczek:1983dx}. }}
for the outcome of these phenomenological DS calculations.
This is 
because the most important for the success of phenomenological DS 
calculations seems the enhancement at somewhat higher values of 
$Q^2$ - e.g., see the humps at $Q^2 \sim 0.4$ to $0.6$
GeV$^2$ in the JM \cite{Jain:qh}
or MRT \cite{Maris:1997tm,Maris:1999nt} $\alpha_{\eff}(Q^2)$, 
dashed curves and dash-dotted curves in Fig. \ref{Fig1}. 
Our $\alpha_{\eff}(Q^2)$ (\ref{ourAlpha_eff}) exhibits such an
enhancement centered around $Q^2 \approx m_A^2/2$, as shown by the 
solid curve representing it in Fig. \ref{Fig1}. This enhancement 
is readily understood when one notices that 
Eq. (\ref{ourAlpha_eff}) has four poles in the complex $Q^2$ plane, 
given by
\begin{eqnarray}
(Q^2)_{1,2} &=& \frac{1}{2}
\left( \,
m_A^2
\mp i\sqrt{4 C_G - m_A^4}
\, \right) \qquad [{\rm poles \, of} \, G^{\Npert}(Q^2)]~,
\label{poles12}
\\
(Q^2)_{3,4} &=& \frac{1}{2}
\left( \, 
-m_A^2
\mp i\sqrt{4 C_A - m_A^4}
\, \right) \qquad [{\rm poles \, of} \, Z^{\Npert}(Q^2)]~.
\label{poles34}
\end{eqnarray}
For $\min\{ C_G, C_A\} > m_A^4/4$ there is no pole on the real axis, 
but a saddle point in the middle of two complex conjugated poles. For 
the DS studies, which are almost exclusively carried out in Euclidean
space, spacelike {$k^2$ (i.e., $Q^2 > 0$ in our convention)} is the 
relevant domain and is thus pictured in Fig. \ref{Fig1}. 
There, the maximum of 
$\alpha_{\eff}(Q^2)$ (\ref{ourAlpha_eff}) at the real axis is at 
$Q^2 \approx m_A^2/2$, i.e., the real part of
its {\it double} poles $(Q^2)_{1,2}$. The height and the width of 
the peak is influenced by both $C_G$ and 
$m_A$. The enhancement of 
$\alpha_{\eff}(Q^2)$ (\ref{ourAlpha_eff}) is thus crucially
determined by the $\langle A^2 \rangle$ condensate through
Eq. (\ref{gluonMass}), and by the manner this condensate contributes 
to the ghost renormalization function, which enters squared
into the effective coupling (\ref{ourAlpha_eff}).

\section{Phenomenology with the condensate-enhanced coupling}
\label{alpha_effPhenomenology}

\noindent
We solved the DS equations for quark propagators and BS equations for 
pseudoscalar $q\bar q$ ($q=u,d,s$) bound states in the same way as in 
our previous phenomenological DS studies 
\cite{Kekez:1996az,Klabucar:1997zi,Kekez:1998xr,Kekez:1998rw}. 
This essentially means as in the JM approach \cite{Jain:qh},
except that instead of JM's $\alpha_{\eff}(Q^2)$, 
Eq. (\ref{ourAlpha_eff}) is employed in the RLA interaction (\ref{RLAkernel}). 
We can thus immediately present the results because we can refer to 
Refs.  \cite{Kekez:1996az,Klabucar:1997zi,Kekez:1998xr,Kekez:1998rw} 
for all calculational details, such as procedures 
for solving DS and BS equations, all model details, 
as well as expressions for inputs such as the aforementioned 
IR-regularized $\alpha_{\pert}(Q^2)$ and explicit expressions for 
calculated quantities, e.g., for $f_\pi$.

\subsection{In the chiral limit}

In the chiral limit, where the bare (and current) quark masses 
vanish, the only parameters are those defining our $\alpha_{\eff}(Q^2)$ 
(\ref{ourAlpha_eff}), namely $m_A, C_A$ and $C_G$. 
It turns out that the initial estimates (\ref{mAestimate}) and 
(\ref{CGCAestimate}), motivated above,
%, $m_A = 0.845$ GeV and $C_G = C_A = (0.640 \,\, {\rm GeV})^4$,
need only a slight modification to provide a very good description
of the light pseudoscalar sector: it is enough to increase the 
estimate $m_A = 0.845$ GeV by just 5\%. That is, the parameter
set
\begin{equation}
C_A = (0.640 \,\, {\rm GeV})^4 = C_G \quad , \quad 
m_A = 0.884 \,\, {\rm GeV} 
\label{StandardParameterSet-old}
\end{equation}
leads to (to begin with) an excellent description of D$\chi$SB,
which gives rise to Goldstone bosons which are also massless 
pseudoscalar $q\bar q$ bound states.
This is seen in the first line of Table \ref{tab:GMOR}: 
our good chiral limit values of the pion decay constant 
($f_\pi \approx 88 \,\, {\rm MeV}$) 
and the $\bar{q}q$ condensate 
[$\langle \bar{q}q \rangle \approx (-214 \,\, {\rm GeV})^3$]
satisfy the Gell-Mann-Oakes-Renner (GMOR) relation (two last 
columns in Table \ref{tab:GMOR}) very well, at the level of 
a couple of percent. These chiral-limit results are similar to, 
e.g., the corresponding results with JM $\alpha_{\eff}(Q^2)$, 
which are also given in Table \ref{tab:GMOR} (in the last line)
for comparison.

\begin{table}
\begin{center}
\begin{tabular}{|c|l|c|c|c|}
\hline
$\alpha_{\eff}$, $C_G$,
 & $\langle\bar{q}q\rangle$ [$\GeV^3$]
 & $f_\pi$ [\GeV]
 & $-\frac{\langle\bar{q}q\rangle}{f_\pi^2}$ [\GeV]
 & $\lim_{m\rightarrow 0} \frac{M_{\pi}^2}{2m}$
     [\GeV] \\
$C_A$, $m_A$
 &
 &
 &
 & \cr
\hline
Eqs. (\ref{ourAlpha_eff}),(\ref{StandardParameterSet-old})
& $ ~(-0.214)^3$ & 0.0882 & 1.261 & 1.293 \\
\hline
Eqs. (\ref{ourAlpha_eff}),(\ref{StandardParameterSet-new})
& $ ~(-0.217)^3$ & 0.0905 & 1.241 & 1.289 \\
\hline
JM $\alpha_{\eff}$ \cite{Jain:qh}
& $ ~(-0.227)^3$  & 0.0898 & 1.368 & 1.401 \\
\hline
\end{tabular}
\end{center}
\caption{The chiral-limit results for $f_\pi$ and $\langle\bar{q}q\rangle$
and the test of the GMOR relation for our $\alpha_{\eff}$ (\ref{ourAlpha_eff})
and the JM one \cite{Jain:qh}.
The quark condensate and the {\em current} quark mass $m$ are calculated at
the renormalization scale $\mu=1~\GeV$.
In DS approach, good values of $f_\pi$ automatically lead to
good description of $\pi^0\to\gamma\gamma$, since the empirically
successful amplitude $T_{\pi^0}^{\gamma\gamma}=1/4\pi^2f_\pi$ is
always obtained analytically in this approach {\it in the chiral
limit} \cite{Maris:1998hc,Roberts:1994hh}.}
\label{tab:GMOR}
\end{table}

The behavior of the momentum-dependent constituent mass function 
$M_q(p^2) \equiv B_q(p^2)/A_q(p^2)$ is also {\it qualitatively} 
similar both to $M_q(p^2)$ found earlier by JM \cite{Jain:qh} and 
ourselves 
\cite{Kekez:1996az,Klabucar:1997zi,Kekez:1998xr,Kekez:1998rw,Kekez:2000aw}
with JM $\alpha_{\eff}$ and to $M_q(p^2)$ obtained now with our 
$\alpha_{\eff}$ (\ref{ourAlpha_eff}) but with different parameters
(this is exemplified by $M_u({-Q^2})$ in Fig. \ref{fig:Mx-lin}). 
Quantitatively, 
for the parameters (\ref{StandardParameterSet-old}) and the 
chiral limit (${\widetilde m}_q = 0$), the constituent quark 
mass $M_q(0) = 0.306$ GeV. This 
is almost 25 \% below both our old results for $M_u(0)$
\cite{Klabucar:1997zi}
obtained with JM $\alpha_{\eff}$ {\it and} our present
$M_u(0)$ in Fig. \ref{fig:Mx-lin}, pertaining to the refitted 
parameters (\ref{StandardParameterSet-new}) and
(much less importantly) to
${\widetilde m}_u \neq 0$ (\ref{newBAREmValues}).
However, the quantitative differences of such a size
are not a problem, since calculations in practice show
that a successful reproduction of the hadronic phenomenology require
just that values of this (anyway unobservable) quantity 
{\it at low} $Q^2$ are of the order of several
hundred MeV, i.e., of the order of typical constituent quark mass,
$M_q(0) \sim M_{\rm nucleon}/3 \sim M_\rho/2$.

The constituent quark mass in the chiral limit, directly 
related to the $\langle\bar{q}q\rangle$ condensate, is also 
very convenient for illustrating the dependence 
of the key D$\chi$SB phenomenon on the model parameters. If we 
vary $C_G$ (for fixed values of $m_A$ and $C_A$) away from its 
phenomenologically favorable value in 
Eq. (\ref{StandardParameterSet-old}), which gives sufficient 
enhancement of $\alpha_{\eff}$, the dynamically generated 
constituent quark mass $M_q(0)$ quickly falls. Beyond some 
critical value of $C_G$, it is always exactly zero, meaning 
that the D$\chi$SB is then completely absent.
The sensitivity of our results to $C_G$ is understandable, 
since from Eqs. (\ref{poles12}) it is clear that $C_G$, in 
combination with $m_A$, influences the height and width 
of the peak of $\alpha_{\eff}(Q^2)$ (\ref{ourAlpha_eff}) 
for spacelike momenta. In spite of this sensitivity, we 
were able to find other combinations of parameter values
which lead to good results. For example, the values 
\begin{equation}
C_A = (0.6060 \,\, {\rm GeV})^4 = C_G \quad , \quad
m_A = 0.8402 \,\, {\rm GeV}
\label{StandardParameterSet-new}
\end{equation}
yield the second line of Table \ref{tab:GMOR}. 
This indicates that there may be an interesting interplay
between $m_A$ and $C_G$ and motivates us to find how the
phenomenologically favorable values of $m_A$ and $C_G$ are
related.  However, we will do it below in 
the more realistic, massive case, away from the chiral limit.
There, the quark bare masses (and the related current masses)
deviate from zero so that empirical masses of pseudoscalar 
mesons can be obtained. 

\begin{figure}
\begin{center}
\epsfxsize = 10 cm \epsfbox{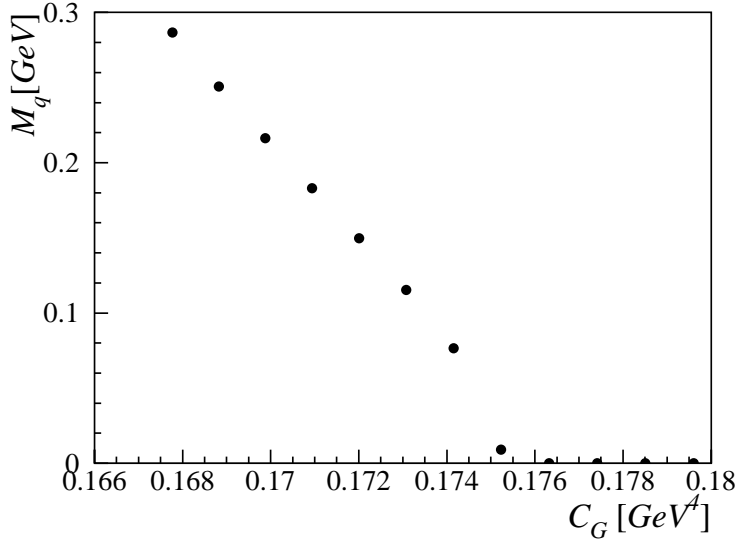}
\end{center}
\caption{The dependence of the dynamically generated 
constituent quark mass $M_q(0)$ on the parameter 
$C_G$ illustrates the disappearance of D$\chi$SB
for unfavorable values of $C_G$: when for given
values of $m_A$ and ${\widetilde m}_q$ (here 
$m_A = 0.884$ GeV and ${\widetilde m}_q = 0$) $C_G$ 
deviates from the value that gives sufficient enhancement of
$\alpha_{\eff}$, the dynamically generated mass
$M_q(0)$ quickly falls. Moreover, beyond some critical value of
$C_G$, it is always exactly zero since the D$\chi$SB
phenomenon then completely disappears.}
\label{fig:ConstituentMassVsCghost}
\end{figure}

\begin{table}
\centering
\begin{tabular}{|c|c|c|l|c|}
\hline
$\alpha_{\eff}$, $C_G$, $C_A$,
& $H$ & $M_H$ [MeV] & $f_H$ [MeV] &
$T_{\pi^0}^{\gamma\gamma}$ [MeV$^{-1}$] \\
$m_A$, ${\widetilde m}_u$, ${\widetilde m}_s$
 &
 &
 &
 & \cr
\hline
Eqs. (\ref{ourAlpha_eff}),(\ref{StandardParameterSet-old})
& $\pi$        & 136.70 & 91.2 & $0.272 \times 10^{-3}$  \\
and (\ref{BAREmValues}) & $K^{+}$      & 520.72 & 112.1  &           \\
\hline
Eqs. (\ref{ourAlpha_eff}),(\ref{StandardParameterSet-new})
& $\pi$        & 136.17 & 93.0 & $0.256 \times 10^{-3}$  \\
and (\ref{BAREmValues}) & $K^{+}$      & 516.28 & 112.5  &           \\
\hline
Eqs. (\ref{ourAlpha_eff}),(\ref{StandardParameterSet-new})
& $\pi$        & 134.96 & 92.9 & $0.256 \times 10^{-3}$ \\
and (\ref{newBAREmValues}) & $K^{+}$      & 494.92 & 111.5  &           \\
\hline
\hline
\hline
{\rm experimental} & $\pi^0$        & $134.9766\pm 0.0006$ & $91.9\pm 3.5$ &
                                $ (0.274 \pm 0.010) \times 10^{-3} $  \\
{\rm values} & $K^{+}$      & $493.677\pm 0.016$ & $112.8\pm 1.0$  &     \\
\hline
\end{tabular}
\caption{The masses and decay constants of pions and kaons, and the 
$\pi^0\to\gamma\gamma$ decay amplitude $T_{\pi^0}^{\gamma\gamma}$, 
obtained in DS
approach with our $\alpha_{\eff}(Q^2)$ (\ref{ourAlpha_eff}).
The first two lines result from the initial parameters
$m_A, C_{A,G}$ (\ref{StandardParameterSet-old}) and the 
quark bare mass parameters (\ref{BAREmValues})
fixed already by the broad JM phenomenological fit \cite{Jain:qh}.
These masses (\ref{BAREmValues}) with another ($m_A, C_{A,G}$)
parameter set (\ref{StandardParameterSet-new})
give the third and the fourth line. Similarly, the fifth and 
the sixth line result from $\alpha_{\eff}(Q^2)$
with $m_A, C_{A,G}$ given by Eq. (\ref{StandardParameterSet-new}),
and the slightly altered bare masses (\ref{newBAREmValues}).
The last two lines are the corresponding experimental values.}
\label{tab:massiveResults}
\end{table}

\subsection{Away from the chiral limit}

We start by noting that both of the two
sets of $(m_A, C_A, C_G)$ values quoted 
above as successful in the chiral limit,
Eqs. (\ref{StandardParameterSet-old}) and 
(\ref{StandardParameterSet-new}), gives a good fit
also away from the chiral limit.
As the first shot, we adopt without any change 
the explicit breaking of chiral symmetry from JM, 
that is, the bare mass parameters (${\widetilde m}_q$) 
of light quarks ($q=u,d,s$) leading to 
the broad phenomenological fit with {\it their}
$\alpha_{\eff}$ \cite{Jain:qh}, namely
\begin{equation}
{\widetilde m}_u = {\widetilde m}_d 
                 = 3.1\cdot 10^{-3}~\GeV 
\quad , \quad
{\widetilde m}_s = 73\cdot 10^{-3}~\GeV \, \, .
\label{BAREmValues}
\end{equation}
These values of ${\widetilde m}_{u,d}$ lead to an excellent 
description of the pion 
as a quasi-Goldstone boson of D$\chi$SB also in conjunction 
with our $\alpha_{\eff}$ (\ref{ourAlpha_eff}) and Eq.
(\ref{StandardParameterSet-old}), as witnessed by 
the first line in Table \ref{tab:massiveResults}, 
where we predict pion mass, weak decay constant, and 
$\pi^0 \to \gamma\gamma$ amplitude very close to their empirical
values (in the seventh line of Table \ref{tab:massiveResults}). 
For the same reason as in the chiral limit, the results are again 
quite sensitive to changes of $C_G$ but not to $C_A$.
Table \ref{tab:results-Bp} illustrates this relatively weak 
sensitivity to the changes of $C_A$ for the case of the parameter 
set (\ref{StandardParameterSet-new})\&(\ref{newBAREmValues}).
Table \ref{tab:results-Bp} shows one can increase (or decrease) 
$C_A$ by a factor of 2, and the results change little.

\begin{table}
\begin{center}
\begin{tabular}{|c|c|l||c|l|}
\hline
 & \multicolumn{2}{c||}{set A}  & \multicolumn{2}{c|}{set B} \\
\hline
 $H$ & $M_H$ & ~~~~$f_H$ & $M_H$ & ~~~~$f_H$  \\
\hline
$\pi$        & 0.132609 & 0.0877711 & 0.136671 & 0.0955129 \\
$K^{+}$      & 0.490013 & 0.107034  & 0.499146 & 0.113735  \\
$s\bar{s}$   & 0.716287 & 0.131859  & 0.727431 & 0.133728  \\
\hline
\end{tabular}
\end{center}
\caption{This table illustrates rather weak sensitivity to changes 
of $C_A$. The ``set A'' of input parameters is given by Eqs. 
(\ref{newBAREmValues}) and (\ref{StandardParameterSet-new})
with the change $C_A \to 2 C_A$.
The ``set B'' of input parameters is given by Eqs.
with the change $C_A \to C_A/2$.
        Meson masses and meson decay constants are in units of \GeV,
while $s\bar{s}$ stands for the non-physical pseudoscalar $s\bar{s}$
bound state.
        }
\label{tab:results-Bp}
\end{table}

The second line of Table \ref{tab:massiveResults} 
reveals that the parameter set 
(\ref{StandardParameterSet-old})\&(\ref{BAREmValues})
works somewhat less well in the strange sector, as the kaon mass 
is 5\% too high.  
However, a deviation of this size is not worrisome
in the present circumstances where we know that the model 
interaction anyway misses some aspects (such as the 
$Q^2 \to 0$ behavior and non-ladder contributions), and 
where we just want to point out that the $\langle A^2 \rangle$ 
condensate is a possible source of the needed enhancement of
$\alpha_{\eff}(Q^2)$. In fact, the empirical success in the
strange sector is quite reasonable considering that 
we used the standard JM mass parameters \cite{Jain:qh},
(as we did also in 
\cite{Kekez:1996az,Klabucar:1997zi,Kekez:1998xr,Kekez:1998rw})
and no refitting was performed there (although 
$\alpha_{\eff}(Q^2)$ was different).

Nevertheless, it is interesting to see what changes are 
brought by refitting. 
If one for example tries the values of $m_A, C_G$ and $C_A$ 
given by Eq. (\ref{StandardParameterSet-new}) instead of 
Eq. (\ref{StandardParameterSet-old}), one gets the third 
and the fourth line in Table \ref{tab:massiveResults}
instead of, respectively, the first and second line.
Thus, the improvement achieved thereby is not significant,
indicating that we should try changes of the bare quark 
masses ${\widetilde m}_q$. It turns out 
that slight changes of the values (\ref{BAREmValues})
are sufficient to achieve agreement with experiment in
the both non-strange and strange sectors.
For example, the parameter set which gives the 
fifth and sixth lines of Table \ref{tab:massiveResults},
thus reproducing the empirical mass of both $\pi^0$ and $K^+$ 
together with good results for their decay constants and 
$\pi^0\to\gamma\gamma$ amplitude $T_{\pi^0}^{\gamma\gamma}$, is
given by $m_A, C_G$ and $C_A$ from Eq. 
(\ref{StandardParameterSet-new}) and by the bare quark masses
\begin{equation}
{\widetilde m}_u = {\widetilde m}_d
                 = 3.046 \cdot 10^{-3}~\GeV
\quad , \quad
{\widetilde m}_s = 67.70 \cdot 10^{-3}~\GeV \, \, .
\label{newBAREmValues}
\end{equation}
This parameter set, Eqs. (\ref{StandardParameterSet-new}) 
and (\ref{newBAREmValues}), is also the one giving the 
gap equation solutions resulting in the momentum-dependent 
constituent mass function $M_q(-Q^2)$ displayed in 
Fig. \ref{fig:Mx-lin}.

\begin{figure}
\begin{center}
\epsfxsize = 6 cm \epsfbox{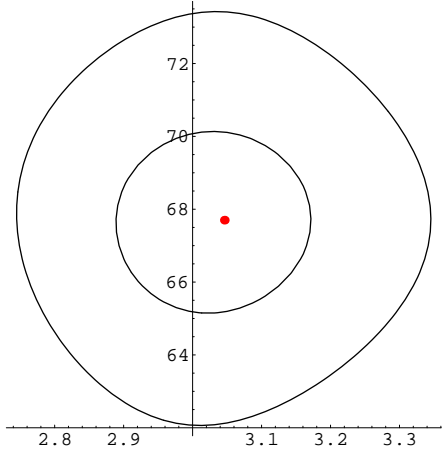}
\end{center}
\caption{The curves are the solutions of the equations $F=2.5\%$ and 
$F=5.0\%$ in $({\widetilde m}_u,{\widetilde m}_s)$ plane. The point
is the position of the simple, non-degenerate minimum at the
%${\widetilde m}_q$-values
bare quark mass values (\ref{newBAREmValues}).
        }
\label{fig:MuMsError}
\end{figure}

The parameter set
(\ref{StandardParameterSet-new})\&(\ref{newBAREmValues})
also gives us a good description of the $\eta$--$\eta'$ complex,
along the lines of our Refs. \cite{Klabucar:1997zi,Kekez:2000aw}.
Although it means employing just a minimal extension of the DS
approach, we must relegate this to another paper \cite{inPreparation}.

The preferred parameter set 
(\ref{StandardParameterSet-new})\&(\ref{newBAREmValues})
is a result of 
a systematic examination of refitting 
possibilities performed by studying the dependence 
on the input parameters
$x=({\widetilde m}_u,{\widetilde m}_s,m_A,C_G,C_A)$
of the function
\begin{equation}
F[x] = \sum_{y}
\left(\frac{y_{\mbox{\rm\scriptsize exp}}-y_{\mbox{\rm\scriptsize th}}}
        {y_{\mbox{\rm\scriptsize exp}}}\right)^2\,\,\times 100\%~,
\label{funct}
\end{equation}
namely the sum of squared differences of the four experimentally 
measured ($y_{\mbox{\rm\scriptsize exp}}$) and presently 
theoretically calculated ($y_{\mbox{\rm\scriptsize th}}$)
quantities $y\in\{ M_{\pi^0}, f_{\pi^\pm}, M_{K^0}, f_{K^\pm} \}$.
We kept choosing $C_{A} = C_{G}$ for simplicity,
since we find that moderate variations of $C_{A}$
do not affect our results much anyway, as already
illustrated by Table \ref{tab:results-Bp}.

\begin{figure}
\begin{center}
\epsfxsize = 12 cm \epsfbox{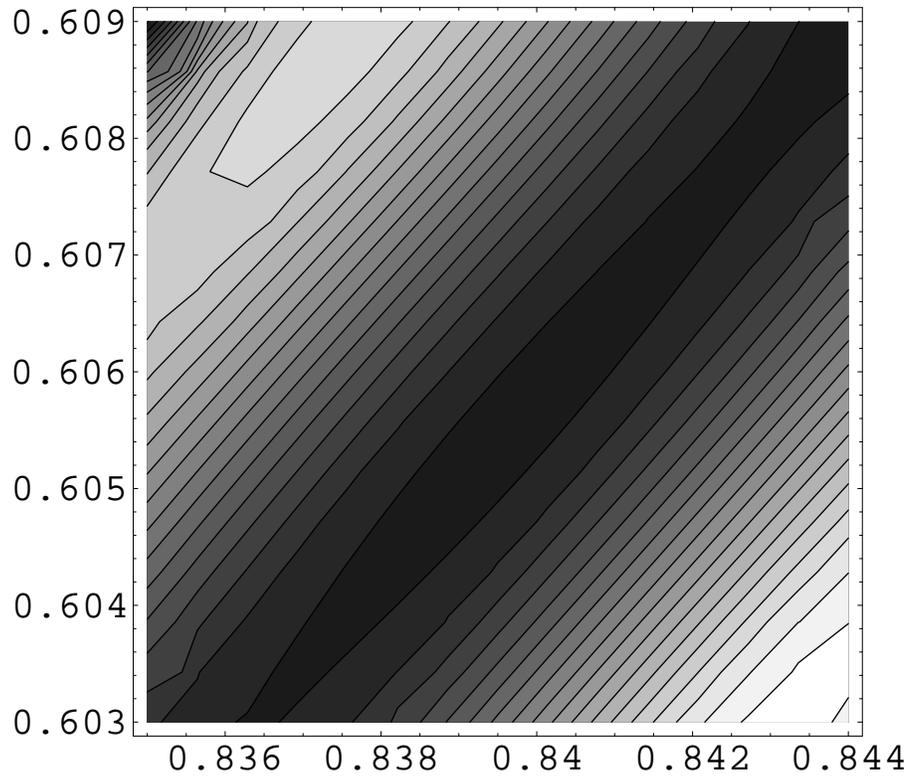}
\end{center}
\caption{$F$ vs. $(m_A,C_G^{1/4})$ contour plot. The darkest
color corresponds to $F\sim 1.5\%$, defining the valley of the
minimal $F$. Conversely, the lighter the shade of gray, the
larger the value of $F$, i.e., the overall difference between
the calculated and experimental quantities.}
\label{fig:MgluonCghostErrorContourPlot}
\end{figure}

Minimization of Eq. (\ref{funct}) shows different respective 
characters of the $\alpha_{\eff}$ parameters $(m_A, C_G, C_A)$
and the mass parameters $({\widetilde m}_u,{\widetilde m}_s)$. 
The point (\ref{newBAREmValues}) in the parameter subspace
$({\widetilde m}_u,{\widetilde m}_s)$ is the location of
a non-degenerate minimum of the function (\ref{funct}).
Thus, the possible values of the bare quark masses 
$({\widetilde m}_u,{\widetilde m}_s)$ can be precisely
restricted by demanding that the function (\ref{funct})
be below certain value. 
Figure~\ref{fig:MuMsError} shows $F=5.0\%$ and $F=2.5\%$ curves 
in the $({\widetilde m}_u,{\widetilde m}_s)$ plane, with $m_A$ 
and $C_G (= C_A)$ fixed at Eq. (\ref{StandardParameterSet-new}). 
At the minimum, for $({\widetilde m}_u,{\widetilde m}_s)$
values (\ref{newBAREmValues}), we obtain $F\approx 1.5\%$.

In contrast to the bare quark masses 
$({\widetilde m}_u,{\widetilde m}_s)$, the parameters 
defining $\alpha_{\eff}$ cannot be determined so unambiguously.
By this we do not mean just the aforementioned weak sensitivity
to $C_{A}$. 
They also cannot be fixed by minimization of $F$ (\ref{funct})
in the same sense as the bare quark masses even though the 
results are very sensitive to $m_A$ and $C_{G}$. The point is 
that $F$ has no simple minimum in the $(m_A,C_G^{1/4})$--plane
as it has in $({\widetilde m}_u,{\widetilde m}_s)$ plane:
Fig. \ref{fig:MgluonCghostErrorContourPlot} 
reveals a minimum in the form of a narrow, straight ``valley'' 
described very well by a linear relation between $m_A$ and 
$C_G^{1/4}$. Thus, in spite of high sensitivity to $m_A$ and 
$C_{G}$, there are many pairs of these quantities which give 
a fit comparable (within few percent) to that resulting from 
the values (\ref{StandardParameterSet-new}), as long as they 
approximately satisfy the linear relation 
\begin{equation}
( C_G )^{1/4} = 0.7742 \, m_A - 0.0444 \,\, {\rm GeV} \,\, .
\label{pravac}
\end{equation}
That is, the function (\ref{funct}) measuring the difference 
between the calculated and experimental values of 
$M_{\pi^0}, f_{\pi^\pm}, M_{K^0}, f_{K^\pm}$ 
has a degenerate minimum 
in the shape of a narrow valley. It is bounded by the values
$(C_G)_{\mbox{\rm\scriptsize min}} \approx (0.6$ GeV$)^4$
and $(C_G)_{\mbox{\rm\scriptsize max}} \approx (0.9$
GeV$)^4$ in the sense that between these values we managed to 
find solutions providing excellent fits ($F$ of the order 1.5\%)
to the empirical values.

\section{Conclusion}
\label{Conclusion}

\noindent
The dimension 2 gluon condensate $\langle A^2 \rangle$ enabled 
the derivation \cite{Kekez:2003ri} of a suitably enhanced 
$\alpha_{\eff}(Q^2)$. 
This effective interaction leads to the sufficiently strong D$\chi$SB 
and successful phenomenology at least in the light sector of 
pseudoscalar mesons.  This opens the possibility
that instead of modeling $\alpha_{\eff}(Q^2)$,
its enhancement at intermediate $Q^2$ may be
understood in terms of gluon condensates, which
seem to provide an important mechanism proposed and studied
for the first time in our recent Ref. \cite{Kekez:2003ri}.
The systematic examination of the parameter space, i.e.,
various fitting possibilities 
set forth in the present paper, allows us to conclude that 
this scenario is compatible with reasonable values of both 
$\langle A^2 \rangle$-condensate and the gauge-invariant dimension 4
gluon condensate $\langle F^2 \rangle$ \cite{Shifman:bx}.
In the relevant momentum region, $\alpha_{\eff}(Q^2)$ (and thus also 
the solutions of DS and BS equations and results for calculated 
measurable quantities) depend only very weakly on $C_A$, which 
parametrizes contributions of dimension 4 condensates to the gluon
propagator.
The essential parameters $C_G$ and $m_A$, on which the dependence
is very strong, are not independent. Thus, due to the relation 
(\ref{pravac}), Eq. (\ref{ourAlpha_eff}) is an essentially 
one-parameter model for $\alpha_{\eff}$, albeit on a relatively 
small interval of $C_G$. 
This can be interpreted as another instance that what counts
is the integrated strength of the interaction. Over the possible
range, we have a continuous set of parameter pairs $(m_A,C_G)$;
their values are such that they give higher peaks at smaller
squared momenta, resulting in similar integrated strengths. 
We find that the 
phenomenologically allowed range of values of the dynamically 
generated gluon mass $m_A$ is in agreement with the lattice 
results \cite{Boucaud:2000nd} on $\langle A^2 \rangle$ in 
the Landau gauge. 
Also, phenomenologically allowed values of $C_G$, which 
parametrizes contributions of dimension 4 condensates to the ghost
propagator, are such that they might be a sign that $C_G$ 
is indeed mostly determined by the dimension 4 
gluon condensate $\langle F^2 \rangle$ \cite{Shifman:bx}.

%%%%%%%%%%%%%%%%%%%%%%%%%%%%%%%%%%%%%%%%%%%%%%%%%%%%%%%%%%%%%%%%%%%

\section*{Acknowledgments}
\noindent The authors gratefully acknowledge the support of 
the Croatian Ministry of Science and Technology contracts 
0119261 and 0098011.

\end{document}